\documentstyle[11pt,psfig]{article}
\def\be{\begin{eqnarray}}
\def\ee{\end{eqnarray}}
\def\bea{\be}
\def\eea{\ee}
\def\calO{{\cal O}}
\def\calB{{\cal B}}
\newcommand{\e}{{\mbox{e}}}
\def\del{\partial}
\def\vr{{\vec r}}

\def\vp{{\vec p}}
\def\vP{{\vec P}}

\def\vs{{\vec \sigma}}
\def\vJ{{\vec J}}

\def\hatr{{\hat r}}
\def\hatk{{\hat k}}
\def\roughly#1{\mathrel{\raise.3ex\hbox{$#1$\kern-.75em%
\lower1ex\hbox{$\sim$}}}}
\def\lsim{\roughly<}

\def\fm{{\mbox{fm}}}
\def\vx{{\vec x}}
\def\EM{{\rm EM}}

\def\zz{{z \bar z}}

\def\abs#1{{\left| #1 \right|}}

\def\nlo#1{{\mbox{N$^{#1}$LO}}}
\def\MS{{\mbox{M1V}}}
\def\mut{{\mbox{M1S}}}
\def\Qt{{\mbox{E2S}}}
\def\rM{{\cal R}_{\rm M1}}\def\rE{{\cal R}_{\rm E2}}

\def\J#1#2#3#4{ {#1} {\bf #2} (#4) {#3}. }
\def\PRL{Phys. Rev. Lett.}

\def\PLB{Phys. Lett. B}

\def\NPA{Nucl. Phys. A}
\def\NPB{Nucl. Phys. B}

\def\PRC{Phys. Rev. C}

\begin{document}

\renewcommand{\thefootnote}{\arabic{footnote}}
\setcounter{footnote}{0}

\vskip 0.4cm
\hfill {\bf KIAS-P99035}

\hfill {\today}
\vskip 1cm

\begin{center}
{\LARGE Effective Field Theory Approach To}
\vskip 0.4cm

{\LARGE $\vec n + \vec p \rightarrow d + \gamma$ At Threshold}

\date{\today}

\vskip 1cm
{\large
Tae-Sun Park$^{a,b,}$\footnote{E-mail: park@alpha02.triumf.ca},
Kuniharu Kubodera$^{a,c,}$\footnote{E-mail: kubodera@sc.edu},
Dong-Pil Min$^{a,d,}$\footnote{E-mail: dpmin@phya.snu.ac.kr}
}

{\large and}

{\large
Mannque Rho$^{a,e,}$\footnote{E-mail: rho@spht.saclay.cea.fr}
}

\end{center}

\vskip 0.5cm

\begin{center}

$^a$
{\it School of Physics, Korea Institute for Advanced Study,
Seoul 130-012, Korea}

$^b$
{\it Theory Group, TRIUMF, 
   Vancouver, B.C., Canada V6T 2A3}

$^c$
{\it Department of Physics and Astronomy, University of South Carolina\\
Columbia, SC 29208, USA\\}

$^d$
{\it Department of Physics, Seoul National University,
Seoul 151-742, Korea}

$^e$
{\it Service de Physique Th\'eorique, CE Saclay,
91191 Gif-sur-Yvette, France}

\end{center}

\vskip 0.5cm

\begin{abstract}
Previously, in an effective field theory formulated by us, 
we have carried out parameter-free calculations 
of a large number of low-energy two-nucleon properties.
An experiment at the Institut Laue-Langevin 
is currently measuring spin-dependent effects 
in the polarized $np$ capture process
$\vec{n}+\vec{p}\rightarrow d +\gamma$ at threshold.
Noting that spin-dependent observables for this reaction 
are sensitive to terms of chiral orders 
higher than hitherto studied,
we extend our effective theory approach
to this process and make {\it parameter-free predictions} 
on the spin-dependent observables. 
\end{abstract}

\newpage

\renewcommand{\thefootnote}{\#\arabic{footnote}}
\setcounter{footnote}{0}

\section{Introduction}\label{intro}
\indent\indent
An experiment by M\"uller {\it et al}~\cite{mueller}
at the Institut Laue-Langevin (ILL) 
is currently measuring spin-dependent effects 
in the thermal neutron capture process
\be
\vec n+\vec p\rightarrow d+\gamma\,,
\label{polnp}
\ee
wherein, as indicated, both the incident neutron
and the target proton are polarized.
At threshold, the initial nuclear state is 
in either the ${}^1S_0 (T=1)$ 
or ${}^3S_1(T=0)$ channel, 
where $T$ is the isospin.
The process (\ref{polnp}) therefore receives contributions 
from the isovector M1 matrix element (M1V)
between the initial ${}^1S_0$ ($T=1$)
and the final deuteron ($T=0$) state,
along with the isoscalar M1 matrix element (\mut) 
and the isoscalar E2 (\Qt) matrix element
between the initial ${}^3S_1$ ($T=0$)
and the final deuteron 
${}^3S_1\!-\!{}^3D_1$ states.\footnote{
The reason why \Qt{} needs to be considered 
will be explained later in the text.
We shall also discuss later the $E1$ contribution
recently considered by Chen, Rupak and Savage \cite{CRS}.}
Since M1V is by far the largest amplitude (see below),
the spin-averaged cross section 
$\sigma_{unpol}(np\rightarrow d\gamma)$
is totally dominated by M1V.
Meanwhile, since the initial ${}^1S_0$ state has $J=0$, 
the M1V cannot yield spin-dependent effects,
whereas \mut{} and \Qt{} can.
This means that, as first noted by Breit and Rustgi\cite{BR71},
the spin-dependent observables in (\ref{polnp})
are sensitive to small isoscalar matrix elements.
To explain the significance of this feature 
in the context of nuclear effective field theory (EFT),
we first give a brief discussion of the 
``chiral filter hypothesis"\cite{KDR}.

The chiral filter hypothesis based on current algebra
was proposed in 1978 prior to the advent of
modern nuclear EFT.
It dictates that corrections to the single-particle
transitions in nuclei for the isovector M1 operator 
and the weak axial charge operator 
should be dominated by one-soft-pion-exchange
two-body currents.
It was shown in \cite{rho91}
that this dominance is naturally explained
by chiral perturbation theory;
the soft-pion exchange diagrams are leading-order
two-body diagrams in chiral counting. 
A thorough nuclear EFT calculation of M1V carried out 
in \cite{pmr} gave a precise, parameter-free
estimate of $\sigma_{unpol}$,
which was in perfect agreement with 
the experimental value \cite{cox};
furthermore, it was confirmed that the major corrections 
to the one-body (single-nucleon) current in M1V
indeed come from one-soft-pion-exchange.
For the axial charge also,
its large enhancement due to the soft-pion exchange 
two-body currents is by now well established \cite{pkmr-int}.   
Thus, for the observables ``protected" by the chiral filter
(i.e., observables essentially determined 
by leading chiral-order terms),
nuclear EFT has been highly successful.

The question now is whether we can profitably 
apply nuclear EFT to observables ``unprotected" 
by the chiral filter.
With leading chiral-order contributions vanishing,
one must consider higher order diagrams,
a task which can quickly become formidable.  
The above-mentioned isoscalar matrix elements, 
M1S and E2S, are examples of the ``unprotected" observable.
As is well known,
the one-body contribution of \mut{} 
is highly suppressed due to the orthogonality
between the initial ${}^3S_1$
and the final deuteron state.
The soft-pion exchange is also suppressed,
there being no isoscalar ${\cal B}^\mu\pi NN$ vertex 
in the leading order chiral Lagrangian, where 
${\cal B}^\mu$ is the external isoscalar field
that couples to the baryonic current.
Due to this double suppression,
the size of $\mut$ becomes even comparable (see below)
to that of \Qt, 
which is a higher-order multipole and
hence, in normal circumstances, can be ignored.
This situation suggests that we must go 
up to an unusually high chiral order 
before getting sensible estimates
of the isoscalar matrix elements  
that govern the spin observables in (\ref{polnp}).
However, we show in this paper (confirming the previous results
reported in \cite{pkmr-int}) that 
nuclear EFT allows us to make
a systematic and reasonably reliable estimation
of M1S and E2S. 

By the very nature of EFT,
an effective Lagrangian involves,
at each order of power counting, 
a certain number of parameters.
These parameters are in
principle calculable from first principles 
for a given scale ({\it e.g.},
lattice QCD~\cite{lepage}) but, in practice, 
they have to be fixed by experimental data.
Once the parameters are so fixed, 
the Lagrangian can then be used to make predictions for
other processes governed by the same parameters. 
However, it often happens that
after fixing the parameters 
there is not much room left for prediction.
A major point of our work is to demonstrate 
that one can make a genuine EFT prediction even 
for the highly suppressed M1S and E2S.
Our {\it bona fide} prediction 
not biased by pre-existing numbers
can be tested soon by the forthcoming 
results from the ILL experiment \cite{mueller}. 

\section{Polarization observables}
\indent\indent
Adopting the convention of M\"uller et al~\cite{mueller}, 
we write the transition amplitude as
\bea
\langle \psi_d, \gamma(\hatk, \lambda) | {\cal T}|
    \psi_{np}\rangle
 = \chi^\dagger_d\, {\cal M}(\hatk, \lambda)\, \chi_{np}
\eea
with
\bea
{\cal M}(\hatk, \lambda) =
 \sqrt{4\pi} \frac{\sqrt{v_n}}{2 \sqrt{\omega} A_s}
 \, \left[
  i (\hatk \times {\hat \epsilon}_\lambda^*)\cdot (\vs_1-\vs_2)\, \MS
\right.
\nonumber \\
 \left.
  - i (\hatk \times {\hat \epsilon}_\lambda^*)\cdot 
  (\vs_1+\vs_2)\,\frac{\mut}{\sqrt{2}}
  + (\vs_1\cdot\hatk \vs_2\cdot {\hat \epsilon}_\lambda^*
   + \vs_2\cdot\hatk \vs_1\cdot {\hat \epsilon}_\lambda^*)
    \frac{\Qt}{\sqrt{2}}
\right]\label{amp}
\eea
where $v_n$ is the velocity of the projectile neutron,
$A_s$ is the deuteron normalization factor
$A_s\simeq 0.8850\ {\rm fm}^{-1/2}$,
and $\chi_d$ ($\chi_{np}$) denotes the spin wave function
of the final deuteron (initial $np$) state.
The emitted photon is characterized by
the unit momentum vector $\hatk$,
the energy $\omega$ and
the helicity $\lambda$,
and its polarization vector is denoted by  
${\hat \epsilon}_\lambda \equiv {\hat \epsilon}_\lambda({\hat k})$.
The amplitudes, $\MS$, $\mut$ and $\Qt$,
represent the isovector M1,
isoscalar M1 and isoscalar E2 contributions, 
respectively.\footnote{These amplitudes are real at threshold.}
These quantities are defined in such a manner 
that they all have the dimension of length,
and the cross section for the unpolarized $np$ system 
takes the form
\bea
\sigma_{unpol}= \abs{\MS}^2 + \abs{\mut}^2 + \abs{\Qt}^2\,.
\label{xsection}
\eea
As we shall see below,  the isoscalar terms
($\abs{\mut}^2$ and $\abs{\Qt}^2$)
are strongly suppressed relative to $\abs{\MS}^2$ 
--- approximately by a factor of $\sim O(10^{-6})$ --- 
so the unpolarized cross section
is practically unaffected by the isoscalar terms.
As mentioned,  
the isovector M1 amplitude was calculated~\cite{pmr} 
very accurately up to ${\cal O} (Q^3)$ 
relative to the single-particle operator.
The result expressed in terms of $\MS$ is:  
$\MS{}=5.78\pm 0.03$ fm,
which should be compared 
to the empirical value
$\sqrt{\sigma_{unpol}^{exp}}= 5.781\pm 0.004$ fm.
In this paper, therefore,
we will focus on the isoscalar amplitudes.

The isoscalar matrix elements are given by
\be
\mut &\equiv& - \frac{\sqrt{2}\omega^{\frac32}}{\sqrt{v_n}} 
 \langle \psi_d^{J_z=1} | \mu^z | \psi_t^{J_z=1}\rangle,
 \nonumber \\
\Qt &\equiv& \frac{\omega^{\frac52}}{\sqrt{8} \sqrt{v_n}} 
  \langle \psi_d^{J_z=1} | Q^{33} | \psi_t^{J_z=1}\rangle
\label{isoscalarME}\ee
with
\be
{\vec \mu}&=& \frac12 \int\! d^3 \vx \, \vx \times \vJ_\EM(\vx),
\nonumber \\
Q^{ij}&=& \int\! d^3 \vx \, (3 x^i x^j - \delta^{ij} \vx^2) J_\EM^0(\vx),
\label{muQdef}\eea
where $J_\EM^\mu(\vx)$ is the EM current.
The spin-triplet initial $np$ and the final deuteron wavefunctions,  
$|\psi_t^{J_z}\rangle$ and $|\psi_d^{J_z}\rangle$,
are given as
\be
\langle \vr | \psi_{t,d}^{J_z}\rangle\equiv
\frac{1}{\sqrt{4\pi} r} \left[
  u_{t,d}(r) + \frac{S_{12}(\hatr)}{\sqrt{8}} w_{t,d}(r) \right]
  \chi_{1 J_z} \zeta_{00}
\ee
with the boundary conditions
\be
\lim_{r\rightarrow \infty} u_d(r) = A_s\, \e^{-\gamma_d r}\;,
\;\;\;\;\;\;\;
\lim_{r\rightarrow \infty} u_{t}(r) = r - a_{t}\;,
\ee
and
\be
\lim_{r\rightarrow \infty} \frac{w_d(r)}{u_d(r)} = \eta_d\;,
\;\;\;\;\;\;\;
\lim_{r\rightarrow \infty} \frac{w_t(r)}{u_t(r)} = 0\;,
\ee
where 
$\gamma_d= \sqrt{B_d m_N}$ with 
$B_d\simeq 2.224575\ \mbox{MeV}$ the binding energy
and $m_N \simeq 939\ \mbox{MeV}$ the nucleon mass;
$\eta_d\simeq 0.0250$ is the asymptotic
$D/S$ ratio of the deuteron;
$a_t \simeq 5.4192\ \fm$ 
is the $^3S_1$ $np$ scattering length,
$S_{12}(\hatr) \equiv 
  3 \vs_1 \cdot \hatr\,\vs_2 \cdot\hatr - \vs_1\cdot\vs_2$,
and $\chi$ ($\zeta$) is the spin (isospin) wavefunctions.

As mentioned, the spin-dependent observables
are sensitive to the isoscalar matrix elements.
Let $I_\lambda(\theta)$ be the angular distribution 
of photons with helicity $\lambda=\pm 1$,
where $\theta$ is the angle between $\hatk$
(direction of photon emission) and a quantization axis of
nucleon polarization.
For polarized neutrons with polarization $\vec{P}_n$ 
incident on unpolarized protons,
the photon circular polarization $P_\gamma$ 
is defined by 
\be
P_\gamma \equiv \frac{I_{+1}(0^\circ) - I_{-1}(0^\circ)}
{I_{+1}(0^\circ) + I_{-1}(0^\circ)}\,.\label{pgammadef} 
\ee
At threshold\footnote{See below for a caveat on the kinematics
of the experiments.}, $P_\gamma$ is given by 
\be         
P_\gamma = |\vP_n|\,
 \frac{\sqrt{2} (\rM - \rE) + \frac12(\rM+\rE)^2}
 {1 + \rM^2 + \rE^2}
\label{pgamma}
\ee
where we have defined the ratios
\bea
{\cal R}_{\rm M1} \equiv \frac{\mut}{\MS}\;,
\ \ \
{\cal R}_{\rm E2} \equiv \frac{\Qt}{\MS}\;.\label{ratio}
\eea
When both protons and neutrons
are polarized (along a common quantization axis)
with polarizations $\vP_p$ and $\vP_n$, respectively,
the anisotropy $\eta_\gamma$ is defined by
\be
\eta_\gamma &\equiv& 
\frac{I(90^\circ) - I(0^\circ)}
{I(90^\circ) + I(0^\circ)}\,, 
\ee
where $I(\theta)= I_{+1}(\theta) + I_{-1}(\theta)$
is the angular distribution of total photon intensity
(regardless of their helicities).  
At threshold, we have
\be
\eta_\gamma &=& pP\, \frac{\rM^2 + \rE^2 - 6 \rM \rE}
     {4(1 - pP) + (4 + pP) (\rM^2 + \rE^2) + 2 pP\, \rM \rE}
\ee
where
\be
pP\equiv \vP_p\!\cdot\!\vP_n\;.
\ee
We note that
$P_\gamma$ is linear in ${\cal{R}}$'s in the leading order.
By contrast, unless $pP\sim 1$,
the anisotropy $\eta_\gamma$ is quadratic
in ${\cal{R}}$'s and hence highly suppressed.
For $pP\sim 1$ (nearly perfect polarization), 
$\eta_\gamma$ becomes ${\cal{O}}(1)$
in powers of ${\cal{R}}$'s
and can be of a substantial magnitude.
This feature suggests the possibility of 
obtaining two independent constraints on ${\cal{R}}$'s,
one from $P_\gamma$ and the other from $\eta_\gamma$, 
which would allow extraction
of the individual values of $\rM$ and $\rE$.
There exist data on $P_\gamma$ due to  
Bazhenov {\it et al.}~\cite{russian},
and the ILL experiment~\cite{mueller} 
is expected to give information on $\eta_\gamma$.

Our argument above is based on the strict threshold kinematics.
A few remarks are in order here.
When the velocity of the projectile neutron $v_n$ goes to zero,
the cross section diverges as $v_n^{-1}$
and consequently the matrix elements as $v_n^{-\frac12}$.
However, the quantity that is actually measured
in experiment is the yield, 
the flux times the cross section.
Since the flux is proportional to $v_n$,
$v_n\sigma$ near threshold is constant with good accuracy,
and one measures the thermal average of 
this $v_n$-independent yield. 
It is customary to translate the measured yield 
into the cross sections using a fixed velocity, 
$v_n=2200$ m/s,
which corresponds to the average neutron velocity
at room temperature.
We adopt this convention throughout this paper,
and we mean by ``at threshold" that 
we are neglecting any higher order
corrections in $v_n$.
Chen et al.{}\cite{CRS} pointed out that at this velocity
the isovector E1 matrix element (we denote it by E1V)
connecting initial $^3P_J$ waves
with the deuteron state can compete with M1S and E2S.
They calculated $P_\gamma$ 
including the E1V contribution
but the corresponding result for $\eta_\gamma$
has not been reported.
The inclusion of E1V affects the asymmetry
$A_{\eta_n}^\gamma(\theta)$ defined and calculated in \cite{CRS}.
However, for photons emitted in the $\theta=0$ direction,
there is no contribution from E1V.
This means that $P_\gamma$,
eq.(\ref{pgammadef}), is not influenced at all 
by the E1V contribution.
Meanwhile, according to \cite{CRS}, 
$\eta_\gamma$ arising from E1V has a very different $pP$ dependence
from that arising from M1S and E2S.
In principle, therefore, 
it should be possible to distinguish experimentally
these two $\eta_\gamma$'s of different origins. 
In the present work limited to the threshold kinematics,
we do not discuss the contribution of E1V.\footnote{
In connection to the above-mentioned thermal averaging, 
we note that the ratio E1V/M1V is proportional to $v_n$,
while $\rM$ and $\rE$ are $v_n$-independent.}

\section{ChPT calculation}
\indent\indent
The chiral counting rules relevant to
electroweak vertices in two-nucleon systems
are well known by now (see, e.g., \cite{pmr}).
The chiral index $\nu$ of a given irreducible diagram
is given by 
\be
\nu = 1 - 2 C + 2 L + \sum_i \nu_i, \ \ \ 
\nu_i = d_i + e_i + \frac{n_i}{2} -2,
\label{naive}\ee
where $C$ is the number of disjoint pieces
($C=2$ for one-body currents and $C=1$ for two-body currents),
and $L$ the number of loops;
$d_i$, $e_i$ and $n_i$ are, respectively, 
the numbers of derivatives/$m_\pi$'s, external fields
and nucleon lines at the $i$-th vertex.
Since chiral symmetry guarantees $\nu_i \ge 0$,
eq.(\ref{naive}) implies that
the one-body EM current $J^\mu_{\rm 1B}$ 
starts at ${\cal O}(Q^{-3})$
while the two-body current $J^\mu_{\rm 2B}$
starts at ${\cal O}(Q^{-1})$.
As spelled out in \cite{pmr}, 
there are additional suppression factors 
of either a kinematical or symmetry origin.
For instance, the space-part of the one-body current
is suppressed by one power in $Q$ relative to its time part.
The detailed counting rules for the two-body isoscalar EM current 
are rather complicated and will be discussed later.

At present, there are essentially two ``alternative" ways of
power counting in setting up an effective field theory
for two-nucleon systems. 
One is the original Weinberg scheme~\cite{weinberg},
in which the leading four-Fermi contact interaction 
and a pion-exchange are treated on the same footing 
in calculating the ``irreducible" graphs 
for a potential that is to be
iterated to all orders in the ``reducible" channel. 
The power counting is done only for irreducible vertices. 
The other is the
``power divergence subtraction" (PDS) scheme~\cite{KSW},
in which only the leading (nonderivative) 
four-Fermi contact interaction is iterated to all orders,
with the higher-order contact interactions
and the pion exchange treated perturbatively.
While the PDS scheme is perhaps more systematic in power counting, 
we believe that the Weinberg scheme is not only consistent 
with the strategy of EFT but
also, in the sense developed below,
more flexible and predictive with possible errors 
due to potential inconsistency 
in power counting generically suppressed.
In our work we adopt the Weinberg scheme.
In this framework, the power counting
for electroweak transition amplitudes
simply reduces to chiral counting 
in the irreducible vertex for the current,
as the current appears only once in the graphs. 
We are then allowed to separate the current matrix element 
into the one-body (single-particle) and
two-body (exchange-current) terms. This framework has the
advantage that it can be straightforwardly applied to n-body
systems for $n>2$. 

Iterating the irreducible vertex 
to all orders in the reducible channel
is equivalent to solving the Schr\"odinger
equation with a corresponding potential. 
This then suggests that we use,
in calculating nuclear responses 
to slowly varying electroweak probes, 
those wave functions computed
with so-called phenomenological ``realistic potentials".
A prime example of these realistic potentials is 
the Argonne $v_{18}$ potential~\cite{argonne} (called in short Av18).
Thus in this approach, referred to as the {\it hybrid approach},
the one-body and two-body transition operators
derived from EFT are sandwiched between the initial and final
two-nucleon wavefunctions distorted by 
a phenomenological N-N potential. 
This procedure of mapping effective field theory 
to realistic wave functions has recently been
justified by means of a cutoff regularization~\cite{PKMR}. 
A similar argument in support of such a hybrid procedure 
has been presented by van Kolck~\cite{vk}. 

Previously we adopted this scheme to study the unpolarized 
$p(n,\gamma)d$ reaction for the thermal neutron.
The cross section for this process was computed 
to next-to-next-to-leading order (NNLO)
in the above-described chiral counting scheme 
for the transition vertex~\cite{pmr}, 
and with the use of the Av18 wave functions.
The resulting theoretical cross section, 
$\sigma_{unpol}^{th}=334\pm 3\ \ {\rm mb}$,
agrees very well with the experimental value
$\sigma_{unpol}^{exp}= 334.2\pm 0.5$ fm \cite{cox};
$\sigma_{unpol}^{th}$ consists of 
the leading 1-body contribution, $305.6\ \ {\rm mb}$, 
and the 2-body exchange-current contribution,
which in conformity with the chiral filter is dominated 
by the soft one-pion exchange.

As shown in \cite{PKMR}, the single-particle matrix
element has negligible uncertainty, 
and therefore the theoretical error
is entirely attributable to the uncertainty 
in the exchange-current operator 
associated with the short-distance part of the interactions.
This short-range behavior is not fully controlled 
in the formulation of \cite{PKMR}. 
The short-distance part introduces 
a scale and renormalization-scheme dependence,
and only those results that are not
sensitive to this uncertainty can be trusted. 
In a framework that uses realistic wave functions, 
the short-distance scale is set 
by a (momentum) cutoff proportional to $r_c^{-1}$,
where $r_c$ is the ``hard-core radius" that removes
the part of the wave function for $r\lsim r_c\neq 0$. 
The net effect of this cutoff is 
that in addition to cutting the radial integrals in
the coordinate space, 
it removes {\it all} zero-range terms in the
current operator including zero-ranged counter terms. 
After the four-fermion counter term 
is removed by the hard core,
there are no more parameters 
in the theory~\footnote{See appendix B
in the second reference of \cite{pmr} 
for the zero-ranged counter term in question. 
In the framework of \cite{KSW}, which avoids
scheme dependence at the expense of predictivity, 
this counter term plays a much more pronounced role,
rendering a {\it bona fide} prediction infeasible~\cite{npKSW}.}. 
This procedure, familiar to nuclear physicists,
may be viewed as a form of exploiting the scheme dependence. 
We shall refer to this procedure as the 
hard-core cutoff scheme (HCCS).
HCCS can be justified for the process
in question by using a (momentum) cutoff $\sim r_c^{-1}$ 
and showing that the counter-term contribution ``killed" 
by the hard core lies within that uncertainty range.\footnote{
Since our calculations are carried out in coordinate space, 
$r_c$ is related to the hard-core radius used 
in standard nuclear physics calculations.}
The $1 \%$ theoretical error assigned 
to $\sigma_{unpol}^{th}$ above
represents uncertainty in this cutoff procedure. 
We will see below that, 
when the chiral filter does not apply, 
the removal of the zero-range
counter terms by the simple hard-core cutoff 
may be much less reliable.

\subsection{One-Body Contributions}
\indent \indent
Either starting from the well-known EM form factors 
of the nucleon,
or carrying out an explicit ChPT calculation
that reproduces the form factors,
we arrive at the following expressions for the one-body operators:
\be
{\vec \mu}_{\rm 1B}(\vr_1, \vr_2) &=& \frac{e}{2 m_N} \sum_{i=1,2}\left[
 \frac{1 + \tau_i^z}{2} \vr_i \times \vp_{i}
 + \frac{\mu_S + \tau_i^z \mu_V}{2} \vs_i \right.
\nonumber \\
&-& \frac{1 + \tau_i^z}{2} 
    \frac{ \left( \vs_i\, \vp_{i}^2 
       - \vp_{i}\,\vs_i\cdot\vp_{i} \right) }{4 m_N^2}
   \left.
   + {\cal O}\left(\frac{\vp^2}{m_N^2}\right)\right],
\nonumber \\
  Q^{ab}_{\rm 1B}(\vr_1, \vr_2) &=& 
  e \sum_{i=1,2}\left[
     \left(3 r_i^a r_i^b - \delta^{ab} r_i^2\right)
     \frac{1+ \tau_i^z}{2} \left(1 + \frac{\vp_i^2}{2 m_N^2}\right)\right.
\nonumber \\
&+& \left. \frac{(2\mu_S-1) + (2\mu_V -1)\tau_i^z}{4 m_N^2}
     U^{ab}_i
+ \calO\left(\frac{\vp^2}{m_N^4}\right) \right]
\ee
with
\be
U^{ab}_i \equiv
\frac32 r_i^a (\vp_{i}\times\vs_i)^b
 + \frac32 r_i^a (\vp_{i}\times\vs_i)^b
  -\delta^{ab} \vr_i \cdot(\vp_{i}\times\vs_i),
\ee
where $\mu_S \simeq 0.87981$ and $\mu_V\simeq 4.70589$.
The matrix elements (\ref{isoscalarME})
can then be obtained by a straightforward calculation.
For example, the leading order 
one-body matrix elements are given by
\be
\mut_{\rm 1B}^{\rm LO} &=&
 - \frac{\sqrt{2} e \omega^{\frac32}}{2\sqrt{v_n} m_N}\,
  \int_0^\infty\!dr\, \left[
       \frac34 w_d(r) w_t(r) +
     \mu_S\, \left(u_d(r) u_t(r) - \frac{w_d(r) w_t(r)}{2}\right)
   \right],
 \nonumber \\
\Qt_{\rm 1B}^{\rm LO} &=&
  \frac{e \omega^{\frac52}}{20 \sqrt{v_n}}\,
 \int_0^\infty\!dr\, r^2 \left[
       \frac{u_d(r) w_t(r) + w_d(r) u_t(r)}{2}
     - \frac{w_d(r) w_t(r)}{\sqrt{8}}
      \right].
\ee
As mentioned earlier, the one-body $\mut$ is suppressed
by the orthogonality 
\be
\int_0^\infty dr\, \left[
 u_d(r) u_t(r) + w_d(r) w_t(r)\right] =0.
\label{ortho}\ee
Using the Av18 wavefunctions,
the numerical results for the sum of the LO and NLO are given as
\be
\mut_{\rm 1B}(\fm) &=& (-4.192 - 0.105) \times 10^{-3}
   = -4.297\times 10^{-3},
   \nonumber \\
\Qt_{\rm 1B}(\fm) &=& (1.401 - 0.007) \times 10^{-3}
     = 1.394 \times 10^{-3}.
\ee
We see that the size of the one-body $\mut$ is 
about $10^3$ times smaller than M1V, 
a suppression which can be traced to the above-mentioned orthogonality
and the smallness of $\mu_S$ as compared to $\mu_V$.

\subsection{Two-body contributions}

\begin{figure}[htbp]
\centerline{\psfig{file=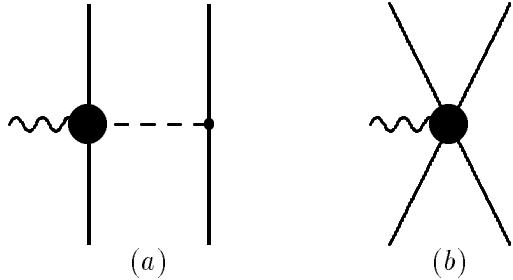}}
\caption[gene]{\protect \small
Generic diagrams for the two-body isoscalar current.
The solid circles include
counter-term insertions and (one-particle irreducible)
loop corrections. The wiggly line stands for the external field
(current) and the dashed line the pion. One-loop corrections for
the pion propagator and the $\pi NN$ vertex need to be
included at the same order.}\label{fig1}
\end{figure}

Generic diagrams for the two-body isoscalar current 
are depicted in Fig.~1.
All the propagators and the vertices
therein should be understood as the renormalized quantities
that include all the loops and counter term insertions
to a given order.
The vertices with non-trivial renormalization
are represented by big filled circles.
The one-loop graphs relevant to the filled circle
in Fig.~$1a$ (Fig.$1b$) are drawn
in Fig.~2 (Fig.~3).

\begin{figure}[htbp]
\centerline{\psfig{file=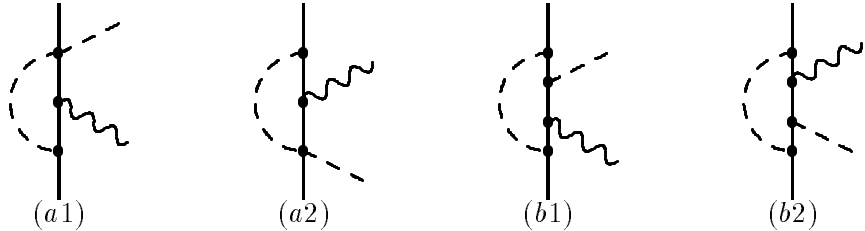}}
\caption[deviate]{\protect \small
One-loop graphs that contribute to the ${\cal B}^\mu \pi NN$ vertex, 
$\calB^\mu$ representing the external isoscalar field.
These diagrams give rise to 
corrections of $\calO (Q^4)$ and higher orders 
to the leading-order one-body term.
\label{fig2} }
\end{figure}

\begin{figure}[htbp]
\centerline{\psfig{file=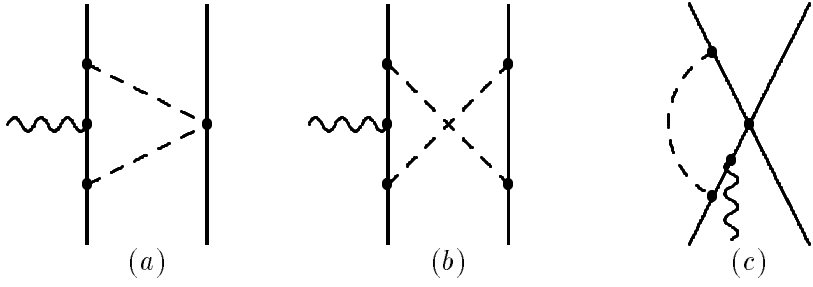}}
\caption[deviate]{\protect \small
One-loop graphs contributing to the two-body baryonic currents.
They come at $\calO (Q^4)$ and higher orders relative to the LO
one-body term. All possible insertions on the external line are
understood.
\label{fig3}
}
\end{figure}

We now proceed to discuss in detail
the counting rules 
for the two-body contributions.
For easy reference, we let \nlo{n} stand for 
contributions of order ${\cal O}(Q^n)$
relative to the leading one-body term.
For the chiral-filter protected M1V,
since the isovector $\gamma\pi NN$ vertex with $\nu_i=0$
enters at the blob of Fig.~1$a$,
the leading correction appears at \nlo{1}
and the one-loops and the counter-terms appear at \nlo{3}.
For the isoscalar current, 
not protected by the chiral filter,
the counting rules are quite different.
The ${\cal B}^\mu\pi NN$ coupling
is highly suppressed by chiral symmetry,
and its first non-vanishing term appears at
\be
\nu_i ({\cal B}^i \pi NN) \ge 2,
\ \ \
\nu_i ({\cal B}^0 \pi NN) \ge 3.
\ee
The same suppression factors appear in the
contact ${\cal B}^\mu NNNN$ vertex.
As a result, leading corrections for the $\mut$ are \nlo{3},
which arise from a chiral Lagrangian~\cite{fettes,pmr} 
of the form
\be
{\cal L}_2 &=& -\frac{g_\rho^2}{8\pi^2 m_\rho^2} 
 \epsilon^{\mu\nu\alpha\beta} 
  (\del_\alpha {\cal B}_\beta - \del_\beta {\cal B}_\alpha)
   \, {\bar B}_v v_\mu\, i \Delta_\nu B_v
\nonumber \\
&-&
i g_4 (\del_\mu {\cal B}_\nu - \del_\nu {\cal B}_\mu)
 {\bar B}_v [S_v^\mu,\, S_v^\nu] B_v\, {\bar B}_v B_v
\ee
where $B_v$ is the nucleon field;
the definitions of the four-velocity $v^\mu$,
the spin-operator $S_v^\mu$ and the covariant
pion-derivative $\Delta^\mu$ can be found in \cite{pmr}.
Although the second term contains an unknown constant
$g_4$, the first term is free from an additional parameter, 
for it is given by the Wess-Zumino term
(related to the Adler-Bell-Jackiw anomaly)
the strength of which can be determined from  
the KSRF relation.
There is also a Lagrangian consisting of
$1/m$ corrections that we shall refer to
as the  ``fixed-term" Lagrangian 
in which all the coefficents are 
uniquely determined \cite{fettes},
\be
{\cal L}^{\rm fixed}
= {\bar B}_v \left[
 \frac{g_A}{m_N} \left\{S_v\cdot D,\, v\cdot \Delta\right\}
 - \frac{g_A}{4 m_N^2}
    (i S_v\cdot \Delta \,D^2 + 2 i S_v\cdot \stackrel
    {\leftarrow}{D}
     \, \Delta \cdot D + \mbox{h.c.})
 + \cdots \right] B_v
\label{fixed}\ee
where $g_A\simeq 1.26$ is the axial coupling constant and
the ellipses denote terms irrelevant to our present study.
The chiral index of the first term is $\nu_i=1$,
while all the other terms have $\nu_i=2$.
It is to be noted, however,
that the $\nu_i=1$ term is proportional to
the energy-transfer,\footnote{
We have replaced the energy transfer
$v\cdot q_i= p_i'^0 - p_i^0$ by
$\frac{{\vec p}_i'^2}{2 m_N} - \frac{\vp_i^2}{2 m_N}$,
where the subscript $i$ is the particle index.}
which is of order of $Q^2/m_N$.
Consequently all the terms in eq.(\ref{fixed})
give \nlo{3} contributions
to the M1 operators.
There is no counter-term contribution to \Qt{}
up to \nlo{4} with which we are concerned.

The loop contributions enter at \nlo{4}
and are finite.
Usually, counter-term contributions enter
-- regardless of whether the loops are divergent or not --
at the same order as the loop contributions.
In our case, however, there is no counter term contribution.
Additional parameters characterizing 
the nuclear potential appear at the 
four-fermion vertex in Fig.~3$c$,
through a Lagrangian of the form
\be
{\cal L}_0 = -\frac12 \sum_A C_A \left({\bar B} \Gamma_A B\right)^2.
\ee
Here $\Gamma_A$'s stand for non-derivative operators,
and $C_A$ are constant coefficients.
{}Due to the Fermi-Dirac statistics,
only two of the four possible $C_A$ are independent.

Summing all the two-body contributions up to \nlo{4}
and Fourier-transforming the resulting expressions
into coordinate space,
we obtain the following isoscalar M1 and E2 operators:
\be
\vec \mu_{\rm 2B}(\vr) &=& \frac{e}{2 m_N} \left[
 \frac14 \left\{{\vec L},\, S_{12}(\hatr)\right\} \tilde C_1^T(r) 
  + \frac12 {\vec L} \tilde C_2(r)
  \right. \nonumber \\
  &&\ 
   + \left(3 \hatr\,\hatr\cdot{\vec S} 
    - {\vec S}\right) \left(\mu_S \tilde C_3^T(r)
     + \frac{r}{12} \tilde C_5^V(r)\right)
     \nonumber \\
     && \left. \ 
     + {\vec S} \left(
      \mu_S \tilde C_4(r) - \frac{1}{6} \tilde C_5^V(r) \right)
       \right]
       + \vec \mu_{1\pi}^{\rm fixed}(\vr),
       \label{mu2Br}\\
       Q^{ij}_{\rm 2B}(\vr) &=& \frac{e}{4} 
        \left(3 r^i r^j - \delta^{ij} r^2\right) \left[
  S_{12}(\hatr) \tilde C_1^T(r) + \tilde C_2(r)\right]
\label{Q2Br}\ee
where 
${\vec S}\equiv \frac12 (\vs_1 + \vs_2)$ and
${\vec L}$ is the orbital angular momentum operator
for the relative motion.
We have separated the contribution from the
fixed-term Lagrangian (\ref{fixed}),
\be
\vec \mu_{1\pi}^{\rm fixed}(\vr) &=& \frac{e}{2 m_N}
 \frac{\tau_1\cdot\tau_2\,m_\pi^3 g_A^2}{16 m_N f_\pi^2}
  \left[
  \left(\frac43 \vec{\cal E}({\vec p})
     - \frac{\vs_1\cdot\vs_2}{3} {\vec L} \right)
      \left[ y_0(m_\pi r) - \delta^3(m_\pi \vr)\right]
  \right.
\nonumber \\
&&\left. +\ \left(3 \vec{\cal E}(\hatr\,\hatr\cdot \vec{p})
  - \frac43 \vec{\cal E}({\vec p})
  -\frac{\left\{S_{12}(\hatr),\,{\vec L}\right\}}{6}
  \right) y_2(m_\pi r)
  \right]
\ee
with
\be
{\cal E}^i(\vec X) = \epsilon^{ijk} r^j 
 \frac{\sigma_1^k \sigma_2^l + \sigma_2^k \sigma_1^l}{2} X^l\,,
\ee
where ${\vec X}$ is either ${\vec p}$ 
or $\hatr\,\hatr\cdot {\vec p}$;
$y_0(x)\equiv \frac{\e^{-x}}{4\pi x}$
and 
$y_2(x)\equiv 
\left(1 + \frac{3}{x} + \frac{3}{x^2}\right)\!y_0(x)$
are Yukawa functions, and
${\vec p}\equiv -\frac{i}{2} (\stackrel{\rightarrow}{\nabla}
  - \stackrel{\leftarrow}{\nabla})$
is the momentum operator with the understanding that
the derivatives act only on wavefunctions.
The $\tilde C$'s are given as
\be
\tilde C_1^T(r) &=&
 \frac{N_1 m_\pi^3}{3} \frac{\e^{-m_\pi r}}{4\pi r}
    \left(1 + \frac{3}{m_\pi r} + \frac{3}{m_\pi^2 r^2}\right)
 + \frac{N_3 m_\pi^3}{24\pi r} \int_0^1 \frac{dz}{(\zz)^2}\,K_3(x),
\nonumber \\
\tilde C_3^T(r) &=&
 \left[\frac{2 m_N}{\mu_S m_\pi} N_{\rm WZ} - N_1\right]
      m_\pi^3 \frac{\e^{-m_\pi r}}{4\pi r}
    \left(1 + \frac{3}{m_\pi r} + \frac{3}{m_\pi^2 r^2}\right)
 - \frac{N_3 m_\pi^3}{8\pi r} \int_0^1 \frac{dz}{(\zz)^2}\,K_3(x)
\nonumber \\
 &&+
 \frac{m_\pi^3}{2\pi r} \int_0^1\frac{dz}{\zz}
   \left[ - \frac{4 N_2}{3} K_3(x) + \frac{N_3}{6} x K_4(x)\right],
\nonumber \\
\tilde C_5^V(r) &=&
  \frac{N_2 m_\pi^3}{\pi r} \int_0^1\frac{dz}{\zz} K_3(x)
\label{C135}\ee
and
\be
{\tilde C}_2(r) &=&
  \vs_1\cdot\vs_2\, \frac{N_1 m_\pi^3}{3} \frac{\e^{-m_\pi r}}{4\pi r}
 - \frac{N_2 m_\pi^3}{\pi r} \int_0^1\frac{dz}{\zz} K_3(x)
 + V_1 \delta(\vr)
\nonumber \\
 &+& \frac{N_3 m_\pi^3}{24\pi r} \left\{
   3 \int_0^1\frac{dz}{\zz} \left[6K_3(x) + 8 K_1(x) + x K_0(x)\right]
   \right.
   \nonumber \\
   &&\left.  -\ (3 + 2 \vs_1\cdot\vs_2)
   \int_0^1\frac{dz}{(\zz)^2} \frac{K_3(x) + 7 K_1(x)}{4} \right\},
\nonumber \\
{\tilde C}_4(r) &=&
  \left[\frac{2 m_N}{\mu_S m_\pi} N_{\rm WZ} - N_1\right]
        m_\pi^3 \frac{\e^{-m_\pi r}}{4\pi r}
 + \frac{N_2 m_\pi^3}{3\pi r} \int_0^1\frac{dz}{\zz} K_3(x)
 + \left[\frac{2 m_N}{\mu_S}(N_{\rm WZ} + g_4) + V_2\right]
      \delta(\vr)
\nonumber \\
 &+& \frac{N_3 m_\pi^3}{24\pi r} \left\{
   \frac12 \int_0^1\frac{dz}{\zz} \left[9K_3(x) + 15 K_1(x) + 2 x K_0(x)\right]
  - 3 \int_0^1\frac{dz}{(\zz)^2} \frac{K_3(x) + 7 K_1(x)}{4} \right\}
  \nonumber \\
  &&
\label{C24}\ee
where $x \equiv \frac{m_\pi r}{\sqrt{\zz}}$, ${\bar z}\equiv (1-z)$
and $K_\nu(x)$ is the irregular modified Bessel functions\footnote{
\protect We follow the definition of the Bessel functions
given in \cite{arfken}.}
of order $\nu$.
The coefficients $N$'s are defined by
\be
N_1 &=& \tau_1\cdot\tau_2 \frac{g_A^4}{128\pi f_\pi^4}\,,
\nonumber \\
N_2 &=& \tau_1\cdot\tau_2 \frac{g_A^2}{128\pi f_\pi^4}\,,
\nonumber \\
N_3 &=& (3 + 2 \tau_1\cdot\tau_2) \frac{g_A^4}{128\pi f_\pi^4}\,,
\nonumber \\
N_{\rm WZ} &=& 
 \tau_1\cdot\tau_2 \frac{g_A g_\rho^2}{48\pi^2 f_\pi^2 m_\rho^2}\,,
\ee
and $V_1$ and $V_2$ are two independent coefficients 
involving $C_A$, 
\be
V_1 &\equiv&
- \vs_1\cdot\vs_2\, \frac{N_1 m_\pi}{3} 
+ \frac{g_A^2 m_\pi}{64\pi f_\pi^2} \sum_A
 \left(\tau_1^b \sigma_1^j C_A\Gamma_A \Gamma_A \tau_1^b\sigma_1^j
  + (1 \leftrightarrow 2)\right),
  \nonumber \\
  (\vs_1+ \vs_2)\,V_2 &\equiv&
   N_1 m_\pi (\vs_1+\vs_2)
   + \frac{g_A^2 m_\pi}{64\pi f_\pi^2} \sum_A
    \left(\tau_1^b \sigma_1^j 
    \left\{C_A\Gamma_A \Gamma_A,\,\sigma_1\right\}
      \tau_1^b\sigma_1^j + (1 \leftrightarrow 2)\right).
\label{V1V2}\ee
In the above equations, 
the terms with $g_4$ and $N_{\rm WZ}$
as well as the fixed-terms 
are due to \nlo{3} contributions,
while all the other terms come from \nlo{4} contributions.
Fortunately, the parameters $V_2$ can be absorbed
into $g_4$,\footnote{
Similarly, $N_{\rm WZ}$ can be renormalized
to absorb some \nlo{4} contribution,
\be
N_{\rm WZ} \rightarrow N_{\rm WZ}' \equiv
 N_{\rm WZ} - \frac{\mu_S m_\pi}{2 m_N} N_1\,.
\label{NWZ}\ee
Unlike the $g_4$ case, however, this
does not reduce the number of unknown parameters, 
since the value of $N_1$ is already known.}
\be
g_4 \rightarrow g_4' \equiv
 g_4 + N_{\rm WZ} + \frac{\mu_S}{2 m_N} V_2\,.
\ee
The $V_1$ term, which in naive counting is \nlo{4},
is further suppressed and plays little role.
The reason is not hard to understand:
$V_1$ appears only in $\tilde C_2(r)$,
which is accompanied 
either by ${\vec L}$ (for \mut) or by $r^2$ (for \Qt).
As a consequence, 
the $V_1$ term contribution has the form
\be
\sim V_1 \,\int d^3\vr\, r^n\, \delta^3(\vr),
\label{V1term}\ee
where $n=4$ (for M1S) and $n=2$ (for E2S),
and eq.(\ref{V1term}) is obviously zero.
Thus we are left with only one parameter $g_4'$.
In the next section we will show 
that we can fix $g_4'$ 
by taking the magnetic moment of the deuteron as input.
This will enable us to make
a totally parameter-free prediction
for the ILL experiment.

\section{Renormalization}
\indent\indent
The two-body contribution to the \mut,
denoted by $\mut_{\rm 2B}$, can be expressed as
\be
\mut_{\rm 2B} = \int_0^\infty\! dr\,
 {\cal F}_t(r) 
- \frac{\sqrt{2} e \omega^{\frac32}}{\sqrt{v_n}}\, g_4'
\int\!d^3\vr\, \frac{u_d(r) u_t(r)}{4\pi r^2}
 \delta^{(3)}(\vr),
\label{naivemut}\ee
where ${\cal F}_t(r)$ consists of
${\tilde C}(r)$'s [see eqs.(\ref{C135}, \ref{C24})]
and the final and initial radial wavefunctions.
The first integral diverges
since ${\cal F}_t(r)$ contains a piece which
behaves like $1/r^2$ for $r\rightarrow 0$.
If the hard-core cutoff scheme (HCCS) defined earlier
can be used, this divergence simply disappears,
as the hard-core kills any contributions
from $r < r_c$ ($r_c$ = hard-core radius).
A hard-core would remove altogether the second integral 
in eq.(\ref{naivemut}).
HCCS was applied quite successfully to
chiral-filter-protected quantities, e.g., M1V~\cite{pmr}.
The numerical results in \cite{pmr} 
exhibit only slight $r_c$-dependence,
a feature that assures the predictivity of HCCS.
For processes not protected by 
the chiral-filter, however, there is no reason
to believe that HCCS continues to be satisfactory.
This is because  
two-body corrections for these processes
come from ``short-ranged physics,"
and therefore imposing a hard-core will throw out the
main part of the relevant physics.
For the process (\ref{polnp})
this problem becomes even more serious 
because the strong suppression of the one-body $\mut$
in our case means that even small changes 
in the two-body contributions can have a
significant influence on the total transition amplitude.
In particular, since we are calculating up to \nlo{4},
it is not justifiable to neglect by fiat
the $g_4'$ term, which is \nlo{3}.

For systematic evaluation of the short-ranged contributions
along with long-ranged contributions,
a transparent method would be to impose a momentum cutoff 
$\Lambda\sim r_c^{-1}$
in performing Fourier transformation.
This scheme would allow us to include 
in a well-defined manner the contributions from $r<r_c$,
whereas they are simply thrown away in HCCS.
Technically, however, it is cumbersome
to introduce a cutoff in Fourier transformation.
Fortunately, for our present purposes,
we can use a convenient scheme equivalent to,
but simpler than, the momentum cutoff method.
For this scheme, we first note that, 
the large-$r$ contributions being scheme-independent,
only the contributions $r \lsim r_c$
involve scheme-dependence.
Secondly, while ${\cal F}_t(r)$ contains
both the $D$-wave and $S$-wave radial functions,
the terms involving the $D$-wave
have higher powers of $r$ 
and hence receive only minor contributions
from the small $r$ region.
Then the only quantity 
that is sensitive to the renormalization scheme
is the small-$r$ contribution of the term containing 
$u_d(r)$ and $u_t(r)$.
For small $r$, however, 
these radial functions can be approximated
as $u_d(r) \simeq u_d'(0) r$ and 
$u_t(r)\simeq u_t'(0) r$.
The small-$r$ contribution ($0 \le r < r_c$)
of the first integral of eq.(\ref{naivemut}) then 
can be written as 
\be
\int_0^{r_c} \!dr\, {\cal F}_t \simeq u_d'(0) u_t'(0) A
\ee
where $A$ is an $r_c$-dependent but wavefunction-independent constant.
This contribution can be absorbed into the redefintion of the parameter
$g_4'$; the resulting $g_4'$ will become $r_c$-dependent.
The net effect of this procedure 
is to remove the $r < r_c$ contribution
while retaining the $r_c$-dependent $g_4'$ term.
Operationally, we can achieve the same goal
by replacing the delta function attached to $g_4'$
with the delta-shell function,
\be
\delta^3(\vr) \rightarrow \frac{\delta(r-r_c)}{4\pi r_c^2}.
\label{deltashell}\ee
For a given value of $r_c$ we can fix $g_4'$
by fitting the deuteron magnetic moment.
We shall call this scheme 
the modified hard-core cutoff scheme (MHCCS).
A regularization scheme similar to MHCCS
has often been used in the literature 
(see, for example, \cite{steele}).
Hereafter we drop the prime on $g_4'$ for simplicity.

In MHCCS, eq.(\ref{naivemut}) is rewritten as
\be
\mut_{\rm 2B} &=& \int_{r_c}^\infty\! dr\,
 {\cal F}_t(r) 
- \frac{\sqrt{2} e \omega^{\frac32}}{\sqrt{v_n}}\, g_4
\frac{u_d(r_c) u_t(r_c)}{4\pi r_c^2}
\equiv
\mut_{\rm 2B}^{\rm finite} + \mut_{\rm 2B}^{\rm zero}.
\label{mut}\ee
We determine $g_4$ through the renormalization
condition for the deuteron magnetic moment,
\be
{\mu_d}_{,{\rm 2B}} = \int_{r_c}^\infty\! dr\,
 {\cal F}_d(r) + 
  2 m_N g_4\, \frac{u_d^2(r_c)}{4\pi r_c^2}
= \mu_d^{\rm exp} - {\mu_d}_{,\rm 1B},
\label{mud}\ee
\be
g_4 = 
 \frac{1}{2 m_N}\,
 \frac{4\pi r_c^2}{u_d^2(r_c)}\,\Delta\mu_d,
\ee
\be
\Delta\mu_d\equiv \mu_d^{\rm exp} - {\mu_d}_{,{\rm 1B}} 
- {\mu_d}_{,{\rm 2B}}^{\rm finite}
\ee
For $r_c$ = 0.01, 0.2, 0.4, 0.6 and 0.8 fm,
we find 
$\Delta \mu_d=$ 0.024, 0.020, 0.022, 0.023 and 0.023
(in units of the nuclear magneton);
the corresponding values of $g_4$ are: 
$g_4(\fm^4)=$ 5.06, 2.24, 0.73, 0.31 and 0.20.
In Table~\ref{T1}
we give the renormalized $\mut$
and its breakdown into the various individual contributions,
for different values of $r_c$ spanning 
a wide range, $0.01\ \fm \le r_c \le 0.8\ \fm$.
\begin{table}[ht]
\begin{center}
\begin{tabular}{|r||r|r|r|r|r|}
\hline
$r_c$ (fm) & 0.01 & 0.2 & 0.4 & 0.6 & 0.8 
\\ \hline
$\mut_{\rm 2B}^{\rm WZ}(\nlo{3})$ &
$-3.384$ & $-3.352$ & $-3.197$ & $-2.789$ & $-2.159$ \\ 
$\mut_{\rm 2B}^{\rm fixed}(\nlo{3})$ &
$-0.938$ & $-0.941$ & $-0.920$ & $-0.823$ & $-0.643$ \\ 
$\mut_{\rm 2B}^{\rm finite}(\nlo{3})$ &
$-4.322$ & $-4.293$ & $-4.117$ & $-3.611$ & $-2.802$ \\ 
$\mut_{\rm 2B}^{\rm finite}(\nlo{4})$ &
$2.658$ & $3.095$ & $2.709$ & $2.049$ & $1.327$ \\ 
$\mut_{\rm 2B}^{\rm finite}$ &
$-1.664$ & $-1.198$ & $-1.408$ & $-1.562$ & $-1.475$ \\ 
\hline
$\mut_{\rm 1B}+ \mut_{\rm 2B}^{\rm finite}$ &
$-5.961$ & $-5.494$ & $-5.704$ & $-5.859$ & $-5.772$ \\
\hline
$\mut$ &
$-2.849$ & $-2.850$ & $-2.852$ & $-2.856$ & $-2.861$ \\
\hline
\end{tabular}
\caption{\label{T1}\protect \small
Individual contributions to \mut\ in
unit of $10^{-3}\ \fm$.
${\mut}_{\rm 2B}^{\rm finite}$ is the sum of
${\mut}_{\rm 2B}^{\rm finite}(\nlo{3})$ and
${\mut}_{\rm 2B}^{\rm finite}(\nlo{4})$,
while
${\mut}_{\rm 2B}^{\rm finite}(\nlo{3})$ is the sum of
the contributions from 
the WZ term, ${\mut}_{\rm 2B}^{\rm WZ}(\nlo{3})$,
and the ``fixed-term," ${\mut}_{\rm 2B}^{\rm fixed}(\nlo{3})$.
We also list the results in the hard-core cutoff scheme (HCCS)
(without the $g_4$ contribution),
$\mut_{\rm 1B}+ \mut_{\rm 2B}^{\rm finite}$,
with $\mut_{\rm 1B}= -4.297\times 10^{-3}\ \fm$.
The last row gives the total \mut{} 
resulting from the MHCCS regularization.}
\end{center}
\end{table}
We see that, while the finite part 
exhibits some $r_c$-dependence, 
the total $\mut$ is completely insensitive 
to the value of $r_c$,
owing to the renormalization
to fit the magnetic moment of the deuteron.
Our results can be summarized as
\be
\mut = (-2.85\pm 0.01)\times 10^{-3}\ \fm\,,
\ee
where the error bar stands for the $r_c$-dependence.
It is highly noteworthy that the $r_c$-independence
holds even when $g_4$ changes by a factor of $\sim\, 26$.

We also observe that the individual contributions 
coming from \nlo{3} and \nlo{4}
are of the same order as the LO one-body contribution.
This feature can be traced to the afore-mentioned facts that
the LO one-body contribution is suppressed
by the orthogonality of the wavefunctions,
and that the \nlo{3} is suppressed due to the
smallness of $g_\rho$.
About half of the \nlo{3} contributions are cancelled
by the \nlo{4} contributions.
In obtaining the above numerical results, 
we used the KSRF value, 
$g_\rho\simeq 5.85$, for the $\rho NN$ coupling constant
featuring in the Wess-Zumino term.
If instead we use a more conventional value, 
$g_\rho\simeq 5.25$,
then the resonance-exchange contribution decreases, 
leading to a larger cancellation between the loop contribution
and the \nlo{3} contributions.
Even with the use of a unrealistically small value of the $g_\rho$
which would cause a complete cancellation 
among the 2-body terms,
we have found 
$\mut = -2.91\times 10^{-3}\ \fm$
(for the entire range of
$0.01\ \fm \le r_c \le 0.8\ \fm$),
which differs only about 2\ \%
from $\mut$ in Table 1.
We note that the MEC contribution reduces 
the one-body contribution
by about 30\ \%.

The two-body contribution to the $\Qt$
turns out to be quite small,
$\Qt_{\rm 2B} = (0.00\pm 0.01)\ \times10^{-3}\ \fm$
for the whole range of $r_c=(0.01 \sim 0.8)\ \fm$.
Combining the one-body and two-body contributions,
we obtain
\be
\Qt = (1.40 \pm 0.01) \times 10^{-3}\ \fm.
\ee
Correspondingly,
the two-body contribution to the deuteron quadrupole moment
also is quite small,  
$Q_{d,{\rm 2B}}= -0.002\ \fm^2$.
This means that the discrepancy 
between the experimental value ($0.286\ \fm^2$)
and the one-body contribution ($0.273\ \fm^2$)
cannot be explained in terms of the two-body effects
as calculated here. 

We now discuss briefly the possibility 
of another renormalization scheme.
Earlier, we invoked eq.(\ref{V1term})
to eliminate the contribution of the $V_1$ term.
However, if we use a non-zero value of $r_c$,
the delta-function gets smeared and therefore
the $V_1$ term can have a finite contribution
even though suppressed by powers of $Q r_c$.
One then might wonder if there exists 
a possibility to renormalize 
the deuteron quadrupole moment $Q_d$ as well as the
magnetic moment by adjusting $V_1$.
This is indeed possible,
and we shall refer to this scheme as MHCCS$^*$.
Since the extension of MHCCS to MHCCS$^*$
is straightforward, we simply quote the results.
For the range of $r_c = (0.01 \sim 0.8)\ \fm$,
we find $\mut= -2.867(1)\times 10^{-3}\ \fm$ and 
$\Qt= 1.348(0)\times 10^{-3}\ \fm$.
Thus MHCCS$^*$ and MHCCS give very similar results.
We have found that, as far as the observables are concerned,
the MHCCS$^*$ results exhibit even a higher degree of
insensitivity to $r_c$ than those of MHCCS.
Meanwhile, the values of $V_1$ and $g_4$ 
in MHCCS$^*$ vary as
$V_1= (3.1\times 10^5 \sim 24.4)\ \fm^3$
and $g_4=(5.05 \sim -0.09)\ \fm^4$,
when $r_c$ changes from 0.01 fm to 0.8 fm.
While neither $V_1$ nor $g_4$ is an observable,
the occurrence of these large coefficients\footnote{
\protect From eqs.(\ref{V1V2}, \ref{NWZ}), 
we can infer that the {\it natural scales} 
of the $V_1$ and $g_4$ are
$N_1 m_\pi \simeq 0.09\ \fm^3$
and $N_{\rm WZ}\simeq 0.03\ \fm^4$, respectively.}
as well as the huge variation of $V_1$ with respect to $r_c$
is rather offending to the {\it naturalness} of the theory.
So we will not consider this scheme seriously.

To illustrate the convergence
of our chiral expansion,
we also present the results
that are renormalized only up to \nlo{3}.
To this order, there are no loops,
and $\Qt$ does not receive any two-particle contribution.
The results for $\mut$ again
turn out to be very close to what we obtained in MHCCS,
exhibiting little $r_c$-dependence:
For $0.01\ \fm \le r_c \le 0.8\ \fm$,
\be
\mut= (-2.84 \pm 0.01)\times 10^{-3}\ \fm\,.
\ee

\section{Discussion}
\indent\indent
The only ``isospin-violating" effect 
that features in the above consideration
is the trivial fact that the two-nucleon system 
can emit both isovector and isoscalar photons.
We examine here whether any significant additional 
isospin-violating effects can arise 
due to higher-order electromagnetic (EM) interactions,
or isospin impurity in the two-nucleon system.
Although in normal circumstances 
these effects give only minor corrections,
they might be important in calculating
highly suppressed amplitudes such as M1S and E2S.
It is particularly important to assess 
those isospin-violating effects in the dominant M1V
which mock the isoscalar amplitude.  
First, we argue that isospin violation
in the initial- and final-state interactions
cannot influence the observables of our concern.
To produce any spin-dependent effects,
the initial $np$ state must have $J\neq 0$,
and hence, at threshold, should be a ${}^3S_1$ state.
If this state could be mixed with ${}^1S_0$
(the only other available state at threshold),
then the dominant M1V transition 
connecting ${}^1S_0$ and the deuteron
would produce a rather significant correction 
to the genuine M1S.
In the two-nucleon systems, however,
the conservation of $J$ dictates that,
even in the presence of isospin-violating interactions,
there be no mixing between the ${}^1S_0$ and ${}^3S_1$ channels.
Next, we need to examine radiative corrections 
to the M1 operator itself.
Instead of considering all possible diagrams,
we study here a typical diagram just to get a rough estimate.
One of the diagrams that give non-vanishing corrections
is obtained from Fig.~3$b$ by replacing 
the one-pion line in it with a photon line.
Let $\xi_{\rm EM}$ be the ratio of the 
contribution of this radiative correction diagram
to that of Fig.~3$b$.
The EM correction becomes more (less) important for
lower (higher) momentum transfers.
When the momentum transfer is of order of $m_\pi$,
we find $\xi_{\rm EM}\sim\alpha/\alpha_\pi\simeq 0.09$,
where 
$\alpha_\pi\equiv 
\frac{g_A^2 m_\pi^2}{16 \pi f_\pi^2} \simeq 0.08$
($f_\pi\simeq 93$ MeV).
Applied to a specific case of $\mut$, however,
this ratio is enhanced by $\mu_V/\mu_S \simeq 5.3$,
as the isovector $\gamma NN$ vertex can contribute to
the isoscalar current due to the isospin breaking EM interaction,
and as a result
the ratio of the EM correction becomes $\sim 1/2$.
Thus the EM correction might be substantial
compared to the \nlo{4}.
We remark, however,
that, in MHCCS (and also in MHCCS$^*$), 
this amount of correction
would cause negligible changes in the final results.

To conclude, 
we recapitulate our predictions in the MHCCS scheme
for the ratios of the relevant matrix elements
and the polarization observables:
\be
\rM&=& (-0.49\pm 0.01) \times 10^{-3},
\nonumber \\
\rE&=& (0.24\pm 0.01) \times 10^{-3},\label{ourpred}
\label{rME}\ee
and strictly at threshold
\be
P_\gamma &=& (-1.04 \pm 0.01)\times 10^{-3},\label{ppred}
\\
\eta_\gamma &=& 0.80\pm 0.02\,,
\label{etais}\ee
where the errors represent the variance
among MHCCS, MHCCS$^*$ and the \nlo{3} MHCCS.
The $r_c$-dependences are negligible.
The above values of $P_\gamma$ and $\eta_\gamma$ are
for complete polarization; that is,
$|\vP_n|=1$ for $P_\gamma$ and $pP=1$ for $\eta_\gamma$.
As noted, 
$\eta_\gamma$ decreases rapidly when $pP$
becomes smaller; for instance,
$\eta_\gamma (pP=0.96) = 0.62\times 10^{-5}$
and
$\eta_\gamma (pP=0.25) = 0.86\times 10^{-7}$.

The results with HCCS are rather different\footnote{These differ
from the results given in \cite{pkmr-int} due to the term
${\mut}_{\rm 2B}^{\rm fixed}(\nlo{3})$ overlooked
in the preliminary calculation reported there.},
\be
\mbox{HCCS}:&&
\rM= (-0.98\pm 0.03) \times 10^{-3},
\\
&&P_\gamma = (-1.73 \pm 0.05)\times 10^{-3},\label{hccs}
\\
&&\eta_\gamma = 0.53\pm 0.01\,,
\label{HCCSres}\ee
where, this time, the errors 
reflect the $r_c$-dependence.
The $\Qt$ remains the same as in eq.(\ref{rME}).

The presently available experimental data for $P_\gamma$ 
\cite{russian} is:
\be
P_\gamma^{exp} = (-1.5 \pm 0.3)\times 10^{-3}\,.\label{pexp}
\ee
Although the prediction in the HCCS, (\ref{hccs}), gives a
$P_\gamma$ closer to (\ref{pexp}), 
we regard our MHCCS result
(\ref{ourpred}) to be more reliable.
The results of Chen, Rupak and Savage \cite{CRS}
for the ratios $\rM$ and $\rE$ computed in the PDS counting scheme
of \cite{KSW} are quite close to ours
(\ref{ourpred}).
Even with the rather generous error estimates 
in \cite{CRS} taken into account, 
this agreement is remarkable.\footnote{
Thus the challenge made in \cite{pkmr-int}
was quickly met by these authors.}

While we believe that MHCCS is more reliable than HCCS, 
we should note that the MHCCS and HCCS results for 
$\rM$ differ by a factor of $\sim 2$.
To be conservative,
we should take this difference as uncertainty
in the treatment of the short-distance physics involved.
($\rE$ does not distinguish 
the two hard-core cutoff schemes.) 
This sort of uncertainty was implied 
in the chiral filter hypothesis of ref.\cite{KDR}.
It has been mentioned in various places 
that, when protected by the chiral filter, 
the HCCS procedure yields quantitatively accurate results.
But the other side of the coin of the chiral filter 
suggests that the HCCS procedure is likely to fail 
when unprotected by the chiral filter,
as in the case of the isoscalar matrix elements at hand.
Our results here indicate
how to understand the possible demise of the 
much used hard-core cutoff procedure, 
a procedure which has long been a standard method
in nuclear physics, in terms of an effective field theory 
regularization.
What comes out as a surprise from this work is that,
despite the uncertainty due to the short-distance behavior, 
one can make a reasonably quantitative calculation
of the observables that are {\it not} protected by
the chiral filter mechanism. 
We await with great eagerness the forthcoming
results of the ILL experiment,
which will bring crucial information on this issue.

\subsection*{Acknowledgments}
\indent\indent
We would like to thank Thomas M. M\"uller for informing us
of his experiment in progress and for useful discussions and
Martin Savage for demonstrating to us the predictive 
power of the effective field
theory approach he and his co-workers have developed. 
We are grateful to C.W. Kim, Director of KIAS, for the 
hospitality and encouragement at KIAS where this work was completed.
The work of KK was partially supported by the NSF Grant
No. PHYS-9602000,
and that of DPM 
by the KOSEF
through CTP of SNU 
and by the Korea Ministry of Education
contract No. 98-2418.

\end{document}